\documentstyle[12pt]{article}
\begin{document}
\begin{center}
{\Large\bf Polygonization and plastic bending of whiskers induced
by pulsed power electron beam}

A.A. Petrova

{\it Tomsk Polytech University, Tomsk, Russia}

\end{center}
\begin{abstract}
Dynamics of bending of NaCl whisker is studied with use of
high speed camera VFU-1 synchronized with high power electron accelerator
GIN-600. The dynamics of bending turns to be non-monotonous and quite
complicated. The figures of etching of whisker after bombardment
by electron beam are analysed, thin polygonization in samples is found.
The time of bending in experiment and time of forming of dislocation walls
computed on the base of theory of plastic bending are compared.
\end{abstract}

It was shown in the paper [1] that intense plastic bending of whiskers
induced by nanosecond pulse of bombardment by high density
electron beam takes place.

Once more process of relaxation of mechanical tensions generated by bombardment
is found. This process represents itself as an intense plastic deformation
connected with arising and sliding of dislocations. This process quite quickly
compensates thermoelastic tensions arising in a sample during bombardment
and decreases resulting tension to be lower than threshold of movement for
cracks thus preventing fragile destruction. However, dynamics and mechanism
of bending were not studied in details.

In this paper estimation of time of plastic bending and its medium speed
obtained from experiment are given and analysis of these data is carried out.

The time  of plastic bending was measured in experiment using methods
described in details in [2].

\vspace*{3mm}

Parameters of experimental system

\begin{tabular}{|l|l|}
\hline
Maximal energy of accelerated electrons  & 0.35 $MeV$\\
\hline
Current of beam of electrons  & 5 $kA$\\
\hline
Density of current  & 4 $kA/cm^2$\\
\hline
Diameter of homogeneous part of beam  & 8 $mm$\\
\hline
Time of pulse &  2-- 30 $ns$\\
\hline
Speed of filming  & 25000-- 25000000 stills per second\\
\hline
Maximal time distinguishing        &     2 $\mu s$\\
\hline
Total number of stills              &         49\\
\hline
Diameter of one still              &               10 $mm$\\
\hline
Vacuum in experimental camera       &        1 $Pa$\\
\hline
\end{tabular}

\vspace*{3mm}

The scheme of experiment is given at Fig. 1. It was carried out in the following
consequence.

A sample of NaCl whisker (6) fixed on special needle holder was
posed in vacuum experimental box (2). The flash lamp (7) for lighting the
sample during filming is established inside of experimental box. The electron
accelerator (1) is connected via vacuum tract with experimental camera (2).
The optical tract connects experimental box with high speed photographic system
VFU-1 which is able to make still-by-still filming of an object.
Synchronization scheme (4) orders moments of switching of accelerator,
flash lamp and high speed photographic camera. The sample brighten by flash
lamp is pierced by high density electron beam during some nanoseconds after
which it begins to bend slowly, and high speed photographic camera carried
out still-by-still filming of the process.

The results of filming are the primary source of information about plastic
bending of whiskers under nanosecond pulse of bombardment by high
density electron beam. The results of experiment are given by Fig. 2, in which
one typical time dependence of deviation h of free end of whisker from
initial position during bending process. One can see that full time of
bending is $\sim 1.5 ms$. There is an induction period 0.1--0.7 ms
during which one bending is small and is not observed practically. The
calculation of thermoelastic tensions arising in thin plates and bars
as a result of inhomogeneous bombardment by high density electron beam
is carried out in [3].

The spectrum of electron beam contains essential number of electrons
with energy no more than 100 keV. Distribution of dose of bombardment
per thickness of whisker is inhomogeneous: as the front part of whisker
is heated more, as rear one is heated less. The thermoelastic tensions arisen
stretch rear part of whisker and compress its front part. Under these tensions
whisker should elastically bend up, i.e. with prominent part turned to beam, and
make elastic bending oscillations with respect to curved quasistatic profile.
Both these effects are of course present but they cannot be seen for photographic
camera. They can be fixed by laser interferometer [4].
Absence of visible deformation during first 0.1 -- 0.7 $ms$ shows also that
mechanical forces acting in crystal are not so large, i.e. thermoelastic
tensions induced by inhomogeneous bombardment are relatively quickly
compensated by tensions of other origin. Those ones must manifest as thermoelastic
tensions are relaxing because of smoothing of temperatures during heat conduction
process.

The characteristic time of smoothing temperature due to heat conductivity is
\begin{equation}
\tau=a^2/\chi (1)
\end{equation}
where $a$ is a characteristic linear scale of temperature
inhomogeneous domain, $\chi$ is a temperature conductivity coefficient.
In our experiment $a$=(0.5 - 1) $b$,
where  $b$ is width of whisker, $\chi= \lambda/(c\rho)$, $\lambda$ is a heat
conductivity coefficient, $c$ is a capacity, $\rho$ is a density.  For NaCl
we have $\lambda=7.4 W/(m\times K)$ at
300 K;  $\rho=2.16\times 10^3 kg/m^3$, $c=0.86\times 10^3 J/(kg\times K)$ hence
$\chi=4\times 10^6 m^2/s$.
Substituting $a=(0.5 - 1.0) b = (48.8 - 97.6)10^{-6} m$ and the value of $\chi$ to
expression (1) we obtain $\tau=(0.6 -- 2.4) ms$.

Hence we conclude that the time of relaxation of quasistatic thermoelastic
tensions generated by inhomogeneous bombardment (and consequently by
inhomogeneous heating) of whisker is about 1 ms. It is easy to find nature of
forces compensating thermoelastic ones. When the value of thermoelastic
tensions stretching upper part of whisker becomes more than value of "flowing
tooth" ($0.7\times 10^8 Pa$ for NaCl [2]), the intense sliding and generation of
dislocations of the same sign begins. These dislocations slide from upper
stretched surface lower to medium unstrained part of a sample and stay on
the threshold of lower compressed part. This is the dislocation mechanism
of bending for plates and bars [4].

As a result of this process upper part of whisker stretched by thermoelastic
tensions is filled by big number of extraplanes limited below by edge
dislocations of one sign. These extraplanes generate tensions preventing from
bending of crystal up, with prominent part turned to beam. As temperature
is smoothed and thermoelastic tensions are relaxed the situation is changed
to opposite one. The upper part of the sample filled with superfluous
extraplanes becomes compressed, and lower one -- stretched. The samples is bent
to opposite side, with concave side turned to beam.

The experiment shows that bending takes place in process of elastic-plastic
non-harmonic decaying oscillations of the sample. Firstly h increases, reaches
maximum, then decreases and again grows up to some stationary value. This is
the typical dynamics of bending (Fig.2).

Two types of pictures are observed. On first ones after bombardment and etching
of the sample one can see branching on lateral sides of whisker turned by
its angle to electron beam (Fig.3). On second ones after bombardment and
etching one can see branching turned by its angle to opposite side (Fig.4).
First type corresponds to whiskers containing occasional defects, cracks or
bubbles, and second one -- to whiskers in those edge dislocations in main
planes of sliding were introduced via stretching on microdeformation machine.
Quantitative pictures (Figs. 3 and 4) are not equivalent but qualitatively
lead to the same result. In both cases the sample bent to the beam.

To understand why figures of etching are different let us turn to mechanisms of
plastic bending and describe behaviour of dislocations during bending.
It is known that during plastic bending dislocations are stayed to be
grouped in sliding lines (see Fig. 5) laying in near sliding plates with
concentrating in most energetically advantageous directions for corresponding
material. Such a direction for
NaCl is $<110>$ (Fig. 6). At Fig. 6 movement of dislocation loop in plane [011]
from bombarded surface to opposite side of crystal in a sample is given.
The lateral (parallel to the vector $b$) sides of loop are spiral dislocations,
and the side of loop which is perpendicular to vector $b$ is an edge dislocation
which reaches surface of crystal faster. If edge dislocation as a component of
dislocation loop does not reach surface of sample one can observe a pyramid-
like pit on lateral surface (100) under etching in point of its crossing with
lateral surface. These pits form strips which one can see at Figs. 3 and 4.
Only after additional heat acting process of decaying of rows of dislocations
in sliding planes and forming of vertical rows of dislocations (polygonization)
can take place (Fig. 7). Let us imagine that these edge dislocations slide
down sliding plates and stand approximately one under another. The picture
of etching is changed but structure of bending of the crystal is not changed
since additional planes which edges are edge dislocations are introduced
from previous direction, e.g. from side opposite to bombardment.

Interaction of edge dislocations of the same sign in parallel sliding planes
in favorable conditions leads to forming of vertical rows of edge dislocations.
The vertical row of dislocations situated on distance $l$ from each other forms
a border of blocks symmetrically turned with respect to each other around axis
parallel to dislocation for an angle $\theta=b/l$.

These processes are most visible during plastic bending of crystal.
Arising of rows of dislocations takes place successively in plastic
deformation process. In our case in NaCl crystals thin polygonization
is observed. It is possible only under essential plastic deformation
when second sliding system begins to act. This sliding plane is perpendicular
to plane (011) on Fig. 6.
Not only pairs of dislocations but also groups of them in orthogonal sliding
planes can be stable. Such a groups in the case of dislocations of the same
sign are so called dislocation walls, i.e. rows of dislocations perpendicular
to sliding planes.
When deformation is inhomogeneous (as in our case) and essential surplus
of dislocations of one sign presents favorable conditions for forming
walls (i.e. rows of dislocations) are established [6].
Reconstruction of border-like dislocation structures from
nets of chaotically situated dislocations, especially after heating,
and in our case electron beam heats the sample creating temperature gradient
about 100 degrees between surface on which electron beam falls and that one
opposite, is connected with decreasing of elastic energy stored in crystal
when dislocations form walls.
Forming of subborders in process of plastic deformation itself takes place
as a result of local reconstruction of dislocations in sliding strips which
also leads to decreasing of elastic energy [7].
If one turns the curved sample by prominent side to beam and continues pulse
bombardment the sample will become straight after several pulse, then it is
curved to opposite side. This phenomenon is agreed with polygonization mechanism
and represents itself as an usual process in ion crystals [7].
In curved samples polygonization etching pits lay along straight lines
corresponding to borders of blocks.

The value of mutual desorientation of blocks is well agreed with number of
dislocations in these borders.
The distance between edge dislocations of the same sign in NaCl wall
$h=100 b$ which is found from pictures of etched samples. Then angle of
desorientation is $\theta=b/100b$ where $b$
is a Burgers vector $b=4\times 10^{-8} cm$ since vertical row of dislocations situated
on distance l from each other forms border of blocks symmetrically turned
with respect to each other around axis parallel to dislocation line,
and the desorientation angle $\theta=b/h$. In our case $\theta=0.6^0$.

Number of dislocation walls calculated on pictures of etched samples is
about 20, so total angle of bending of whisker coincides with theoretical
angle $12^0$.
It is necessary to find speeds of sliding of edge dislocations in NaCl
crystal under action of tensions and to estimate time necessary for forming
dislocation walls. Real tensions acting in NaCl whisker under bombardment by
fast electrons are $\sigma=0.7\times 10^8 Pa$.
In [8] the dependence $v(\sigma)$ was found from which one concludes that speed of
edge dislocations at such tensions is about $10^2 m/s$.
If medium thickness of whisker about 60 $\mu m$ dislocation needs
$6\times 10^{-7}s$ to pass distance from place of arising to opposite side of
crystal. This time is enough for construction of dislocation walls since time
of bending is about 1 ms.

Conclusions.

1. Dynamics of elastic-plastic bending of NaCl whisker induced by nanosecond
pulse of bombardment by intense electron beam is studied with use of high-speed
still-by-still filming. The measured time of irreversible plastic bending of
whisker lies in an interval 1.5 - 2.5 ms and practically coincides with
characteristic time of relaxation of quasistatic thermoelastic tensions arising
due to inhomogeneous heating of whisker by beam.

2. The observed bending is not elastic in all characteristics: time, space and
thermodynamic ones. The main contribution to bending is given by
elastic-plastic deformation connected with movement and multiplication of
dislocations.

3. Different pictures of etching of whiskers bent to beam are found. At
detailed consideration it turned to be that dislocation walls are seen on the
pictures. The estimations of sliding of dislocations are made. The angle of bending
of samples is connected with the angle of desorientation of blocks forming
during polygonization. These estimations are agreed with experimental data.

\unitlength=.5mm
\begin{picture}(150,100)
\thicklines
\put(5,5){\framebox(40,20)}
\put(100,5){\framebox(40,20)}
\put(39,8){\small 5}
\put(130,8){\small 4}
\put(40,50){\line(1,0){40}}
\put(40,70){\line(1,0){40}}
\put(40,50){\line(0,1){20}}
\put(75,52){\small 2}
\put(80,50){\line(0,1){6}}
\put(80,70){\line(0,-1){6}}
\put(44,55){\small 7}
\put(100,50){\line(0,1){6}}
\put(100,70){\line(0,-1){6}}
\put(100,50){\line(1,0){40}}
\put(100,70){\line(1,0){40}}
\put(140,50){\line(0,1){20}}
\put(132,52){\small 3}
\put(50,1){\line(1,0){25}}
\put(50,1){\line(0,1){29}}
\put(75,1){\line(0,1){29}}
\put(71,5){\small 1}
\put(50,30){\line(1,1){10}}
\put(75,30){\line(-1,1){10}}
\put(60,40){\line(1,0){5}}
\put(120,25){\line(0,1){25}}
\put(62,40){\vector(0,1){10}}
\put(63,40){\vector(0,1){10}}
\put(64,40){\vector(0,1){10}}
\put(80,60){\vector(1,0){20}}
\put(52,55){\framebox(10,5){\small 6}}
\put(61,61){\line(1,0){8}}
\put(78,50){\line(0,-1){28}}
\put(100,22){\line(-1,0){22}}
\put(45,65){\circle{5}}
\put(42,25){\line(0,1){25}}
\put(75,8){\line(1,0){25}}
\put(50,-10){Fig.1}
\end{picture}


\unitlength=0.9cm
\begin{picture}(20,7)
\thicklines
\put(6,0){\line(0,1){6}}
\put(6,0){\line(1,0){7}}
\put(6,5){\vector(0,1){1}}
\put(13,0){\vector(1,0){1}}
\put(4.5,5.5){h, mm}
\put(13,-.5){t, ms}
\newcounter{S}
\multiput(6.1,2)(0,2){2}{\line(-1,0){.2}}
\multiput(5.6,2)(0,2){2}{\addtocounter{S}{1}\arabic{S}}
\put(6,.1){\line(0,-1){.2}}
\put(5.7,-.7){0.6}
\put(8,.1){\line(0,-1){.2}}
\put(7.7,-.7){0.8}
\put(10,.1){\line(0,-1){.2}}
\put(9.7,-.7){1.0}
\put(12,.1){\line(0,-1){.2}}
\put(11.7,-.7){1.2}
\special{em:linewidth 1.6pt}
\put(6.87,0){\special{em:point 37}}
\put(7.15,1.0){\special{em:point 38}}
\put(7.42,0.0){\special{em:point 39}}
\put(8.25,0.0){\special{em:point 40}}
\put(8.52,1.0){\special{em:point 41}}
\put(8.80,0.0){\special{em:point 42}}
\put(9.07,1.0){\special{em:point 43}}
\put(9.35,0.0){\special{em:point 44}}
\put(9.62,0.0){\special{em:point 45}}
\put(9.90,2.6){\special{em:point 46}}
\put(10.17,1.4){\special{em:point 47}}
\put(10.45,2.8){\special{em:point 48}}
\put(10.75,2.6){\special{em:point 49}}
\put(11.00,2.8){\special{em:point 50}}
\put(11.27,2.8){\special{em:point 51}}
\put(11.55,0.0){\special{em:point 52}}
\put(11.82,1.0){\special{em:point 53}}
\put(12.10,3.0){\special{em:point 54}}
\put(12.37,4.0){\special{em:point 55}}
\put(12.65,6.0){\special{em:point 56}}
\put(12.92,5.2){\special{em:point 57}}
\put(13.20,6.0){\special{em:point 58}}
\put(13.47,6.0){\special{em:point 59}}
\special{em:linewidth 1.6pt}
\special{em:line 37,38}
\special{em:line 38,39}
\special{em:line 39,40}
\special{em:line 40,41}
\special{em:line 41,42}
\special{em:line 42,43}
\special{em:line 43,44}
\special{em:line 44,45}
\special{em:line 45,46}
\special{em:line 46,47}
\special{em:line 47,48}
\special{em:line 48,49}
\special{em:line 49,50}
\special{em:line 50,51}
\special{em:line 51,52}
\special{em:line 52,53}
\special{em:line 53,54}
\special{em:line 54,55}
\special{em:line 55,56}
\special{em:line 56,57}
\special{em:line 57,58}
\special{em:line 58,59}
\put(5,-1.5){Fig.2}
\end{picture}

\end{document}